\documentclass[showpacs,superscriptaddress,twocolumn,tightenlines,amsmath,amssymb,nofootinbib]{revtex4}

\usepackage{graphicx}
\usepackage{dcolumn}
\usepackage{bm}
\usepackage{amssymb}
\usepackage{amsmath}
\usepackage{epsfig}    
\usepackage{color}
\usepackage{slashed}
\newcolumntype{I}{!{\vrule width 1.3pt}}
\begin{document} 
\title{
First constraint on the mass of doubly-charged Higgs bosons in the same-sign diboson decay scenario at the LHC
}
\preprint{UT-HET 080}

\author{Shinya Kanemura}
\affiliation{Department of Physics, University of Toyama, \\3190 Gofuku, Toyama 930-8555, Japan}
\author{Kei Yagyu}
\affiliation{Department of Physics, National Central University, \\Chungli 32001, Taiwan}
\author{Hiroshi Yokoya}
\affiliation{Department of Physics, University of Toyama, \\3190 Gofuku, Toyama 930-8555, Japan}

\begin{abstract}

When the doubly-charged Higgs bosons $H^{\pm\pm}$ mainly decay into the same-sign dilepton, 
a lower bound on the mass is around 400~GeV by the current LHC data. 
On the other hand, no such bound has been reported by using the data at LEP and at the LHC for the case where 
the same-sign diboson decay $H^{\pm\pm}\to W^{\pm(*)} W^{\pm(*)}$ is dominant. 
We study limits on the mass for such a case by using the current experimental data. 
From the precise measurement of the total width of the Z boson at the LEP experiment, 
the mass below 43 GeV is excluded with the 95\% confidence level.   
It turns out that the results from four charged lepton searches at LEP do not provide 
any significant constraint. 
We show that a new lower bound is obtained in the diboson decay scenario at the LHC with 
the collision energy to be 7 TeV and the integrated luminosity to be 4.7 fb$^{-1}$. 
By using the data of the same-sign dilepton events, 
the lower limit is found to be 60 GeV at the 95\% confidence level. 
By the extrapolation of the data to 20~fb$^{-1}$ with the same collision energy, 
the lower limit is evaluated to be 85 GeV.  

\pacs{12.60.Fr, 14.80.Fd}

\end{abstract}
\maketitle
\newpage
A Higgs boson has been discovered with the mass around 126 GeV in the diphoton and 
the four lepton channels at the LHC~\cite{LHC_Higgs}. 
Observed properties of the boson are consistent with those of the Higgs boson predicted in the standard model (SM). 
However, this fact does not necessarily mean that the Higgs sector is the minimal one assumed in the SM. 
There is no fundamental principle for such a form with an isospin doublet scalar field.  
In fact, extended Higgs sectors with more scalar fields 
are often introduced in new physics models beyond the SM, which are motivated to explain phenomena such as 
neutrino oscillation, the existence of dark matter and the baryon asymmetry of the Universe. 
The SM-like Higgs boson can also appear in these extended Higgs sectors. 
Therefore, by exploring the extra Higgs bosons at collider experiments, 
the structure of the Higgs sector can be determined, and thereby the new physics model can be probed. 

The Higgs triplet model (HTM) is one of the simple but important extended Higgs sectors. 
It can provide the origin of tiny neutrino masses with the type-II seesaw mechanism~\cite{typeII}. 
The Higgs sector is composed of an isospin triplet field ($\Delta$) with hypercharge $Y=1$ and a doublet field ($\Phi$) with $Y=1/2$. 
The generated neutrino masses are proportional to the vacuum expectation value (VEV) of 
the triplet field $v_\Delta$. 
It is well known that $v_\Delta$ is constrained to be much smaller than the VEV of the doublet field $v_\phi$ 
by the electroweak precision data at the LEP/SLC experiments~\cite{LEP_SLC}. 
One of the striking features of the model is the appearance of the doubly- ($H^{\pm\pm}$) and singly-charged ($H^{\pm}$) Higgs bosons 
in addition to two CP-even Higgs bosons ($h$ and $H$) and the CP-odd Higgs boson ($A$). 
The discovery of these particles, especially $H^{\pm\pm}$, is essentially important to identify the model. 

These additional Higgs bosons can be produced by collider experiments if kinematically accessible~\cite{Region1, Han,Nomura,HTM_mass_diff,AKY}. 
The main decay mode of $H^{\pm\pm}$ strongly depends on $v_\Delta$. 
They decay into the same-sign dilepton when $v_\Delta$ is smaller than about 0.1 MeV, 
which can be a clear signature for the discovery. 
Searches for $H^{\pm\pm}$ have been performed at LEP, Tevatron and the LHC. 
Since $H^{\pm\pm}$ have not been discovered yet, lower limits for the mass of $H^{\pm\pm}$ 
are obtained to be 409 GeV, 398 GeV and 375 GeV 
in the case where the branching fraction of $H^{\pm\pm}\to e^\pm e^\pm$, 
$H^{\pm\pm}\to \mu^\pm \mu^\pm$ and $H^{\pm\pm}\to e^\pm \mu^\pm$ are respectively assumed to be 100\%~\cite{400GeV}. 
On the contrary, when $v_\Delta$ is much greater than 0.1 MeV, the decays into diboson 
$H^{\pm\pm}\to W^{\pm(*)} W^{\pm(*)}$ are dominant\footnote{When $H^\pm$ is lighter than $H^{\pm\pm}$, the main decay mode of $H^{\pm\pm}$ is $H^{\pm\pm}\to W^{\pm(*)} W^{\pm(*)}$  or $H^{\pm\pm}\to W^{\pm*} H^\pm$, depending on the mass difference and $v_\Delta$~\cite{HTM_mass_diff,AKY}. }.  
They can be identified by the same-sign dilepton with the missing energy signature via the leptonic decay 
of the W bosons. 
Although this scenario is equivalently important to the case of the dilepton decay from the theoretical point of view~\cite{Han,Nomura}, 
to our knowledge no substantial bound has been reported for the $H^{\pm\pm}$ search via the decay of $H^{\pm\pm}\to W^{\pm(*)} W^{\pm(*)}$. 

In this Letter, 
we discuss limits on the mass for the case of the diboson decay scenario by using the current data at LEP and the LHC. 
First, we consider the constraint from the precise measurement of the total width of the Z boson at the LEP experiment~\cite{PDG}, and 
then we evaluate the lower limit for the mass of $H^{\pm\pm}$ by using the
data of the same-sign dilepton events at the LHC with 
the collision energy to be 7 TeV and the integrated luminosity to be 4.7 fb$^{-1}$~\cite{ATLAS}. 
Finally, we extrapolate the results for the limit with the 20 fb$^{-1}$ data. 

The most general potential in the HTM is given by 
\begin{align}
& V = m^2 \Phi^\dagger \Phi + M^2 {\rm Tr}(\Delta^\dagger \Delta) +
  [ \mu \Phi^T i \tau_2 \Delta^\dagger \Phi + {\rm h.c.} ] \notag\\
& + \lambda_1 (\Phi^\dagger \Phi)^2 + \lambda_2 \left[{\rm
                                                  Tr}(\Delta^\dagger
                                                  \Delta) \right]^2
  + \lambda_3 {\rm Tr}\left[(\Delta^\dagger \Delta)^2\right]\notag\\
&  +\lambda_4 (\Phi^\dagger\Phi) {\rm Tr}(\Delta^\dagger \Delta)
  +\lambda_5 \Phi^\dagger \Delta \Delta^\dagger \Phi. 
 \end{align}  
The scalar fields $\Phi$ and $\Delta$ can be parameterized as 
\begin{align}
\Phi&=\left[
\begin{array}{c}
\phi^+\\
\frac{1}{\sqrt{2}}(\phi+v_\phi+i\chi)
\end{array}\right],\quad \Delta =
\left[
\begin{array}{cc}
\frac{\Delta^+}{\sqrt{2}} & \Delta^{++}\\
\Delta^0 & -\frac{\Delta^+}{\sqrt{2}} 
\end{array}\right], 
\end{align}
with $\Delta^0=(\delta+v_\Delta+i\eta)/\sqrt{2}$. 
The VEVs satisfy $v^2 \equiv v_\phi^2+2v_\Delta^2=(\sqrt{2}G_F)^{-1}=$ (246 GeV)$^2$, where 
$G_F$ is the Fermi constant. 
The mass eigenstates of the physical scalar bosons are obtained by introducing the mixing angles $\alpha$, $\beta$ and $\beta'$, 
and their formulae are expressed in terms of the parameters in the potential~\cite{AKKY_full}. 

The electroweak rho parameter is predicted to deviate from unity in the HTM by $\rho \simeq 1-2v_\Delta^2/v^2$. 
Since the experimental value of the rho parameter is close to unity; i.e., $\rho_{\text{exp}}=1.0004^{+0.0003}_{-0.0004}$~\cite{PDG}, 
$v_\Delta$ must be less than about 3.5 GeV at 95\% confidence level (CL). 
Under $v_\Delta \ll v$, six physical scalar states;  
$H^{\pm\pm}~(=\Delta^{\pm\pm})$, $H^\pm$, $A$ and $H$ are almost composed of the component scalar fields from $\Delta$, 
so that we call them as the triplet-like Higgs bosons. 
On the other hand, $h$ can be regarded as the SM-like Higgs boson. 

The mass difference  between $H^{\pm\pm}$ and $H^{\pm}$ is given by 
\begin{align}
 m_{H^{++}}^2-m_{H^{+}}^2 \simeq m_{H^{+}}^2-m_{H}^2  \simeq -\frac{\lambda_5}{4}v^2, 
\end{align}
with $m_A^2\simeq m_H^2$. 
There are three patterns of the mass spectrum for the triplet-like Higgs bosons. In the case with $\lambda_5=0$, all the triplet-like Higgs bosons are degenerate in mass. 
On the other hand, in the case of $\lambda_5>0$ ($\lambda_5<0$), the mass spectrum is 
$m_{A}>m_{H^+}>m_{H^{++}}$ 
($m_{H^{++}}>m_{H^+}>m_A$). 
We call this mass spectrum as Case~I (Case~II)~\cite{AKY}.

Neutrino masses can be deduced via $v_\Delta$ as~\cite{typeII}
\begin{align}
(\mathcal{M}_\nu)_{ij}=\sqrt{2}h_{ij}v_\Delta,  \label{eq:mn}
\end{align}
where the Yukawa coupling constants $h_{ij}$ are defined by 
\begin{align}
\mathcal{L}_Y&=h_{ij}\overline{L_L^{ic}}i\tau_2\Delta L_L^j+\text{h.c.}  \label{nu_yukawa}
\end{align}
Eq.~(\ref{eq:mn}) tells us that the magnitude of the Yukawa coupling $h_{ij}$ is determined by a fixed value of $v_\Delta$, because 
the value of the left hand side of the equation is given from neutrino data~\cite{PDG}. 

The decay property of $H^{\pm\pm}$ strongly depends on $v_\Delta$ and the mass spectrum of the triplet-like Higgs bosons. 
When we consider the degenerate case or Case~I, 
the dominant decay modes are $H^{\pm\pm}\to \ell^\pm\ell^\pm$ or $H^{\pm\pm}\to W^{\pm (*)} W^{\pm (*)}$ in the case of $v_\Delta\lesssim 0.1$ MeV 
or $v_\Delta\gtrsim 0.1$ MeV, respectively. 
On the other hand, in Case~II, above two main decay modes can be replaced by the cascade decay mode; i.e., 
$H^{\pm\pm}\to W^{\pm *}H^\pm \to W^{\pm *}W^{\pm *}H/A $. 

We here focus on the scenario where $H^{\pm\pm}\to W^{\pm (*)}W^{\pm (*)}$ are dominant, 
so that we consider Case~I or the degenerate case with $v_\Delta\gtrsim 0.1$ MeV. 
We then discuss the bound for $m_{H^{++}}$ by using the current experimental data. 

\begin{center}
\begin{table*}[t]
{\small
\hfill{}
\begin{tabular}{l|c|c|c|c|c|c|c}\hline\hline
$m_{H^{++}}$ [GeV] &40& 50& 60 &70 &80&~~90~~&~~100~~  \\\hline
Width [eV] & 3.58$\times 10^{-4}$& 1.89$\times 10^{-3}$   & 7.70$\times 10^{-3}$ & 2.69$\times 10^{-2}$ & 8.63$\times 10^{-2}$ & 0.320 & 1.86  \\\hline
Basic cut [fb] & 20.9& 16.7& 13.0 & 10.1 & 7.63 & 4.95 & 2.78  \\\hline
$M_{\ell\ell}$ cut [fb] & 16.3 &14.8& 12.2 & 9.69 & 7.45 & 4.87 & 2.74\\\hline\hline
\end{tabular}}
\hfill{}
\caption{Signal cross sections of the process in Eq.~(\ref{sig1}) for the $\mu^+\mu^+$ channel 
for fixed values of $m_{H^{++}}$ in each step of the kinematic cuts. 
The decay width of $H^{\pm\pm}$ is calculated in the case of $v_\Delta=1$ GeV for reference. }
\label{cross}
\end{table*}
\end{center}

The LEP experiment was operated with the electron-positron collision energy to be at the Z boson mass (LEP~I) and to be up to about 209 GeV (LEP~II). 

From the LEP~I experiment, the total width of the Z boson has been precisely measured, and it can be used to 
constrain $m_{H^{++}}$ whose value is smaller than the half of $m_Z$, independently of the decay modes of $H^{++}$. 
The total decay width of the Z boson receives the sizable correction from the partial width for this decay mode as
\begin{align}
\hspace{-3mm}\Gamma_{Z\to H^{++}H^{--}}=\frac{G_Fm_Z^3}{6\pi\sqrt{2}}(1-2s_W^2)^2\left(1-\frac{4m_{H^{++}}^2}{m_Z^2}\right)^{\frac{3}{2}}, \notag
\end{align}
where $s_W$ is the sine of the weak mixing angle. 
Using the current experimental data for the Z boson width, $\Gamma_Z(\rm exp)=2.4952\pm0.0023$~GeV, and the SM prediction,
$\Gamma_Z(\rm SM)=2.4960\pm0.0002$~GeV~\cite{PDG}, we
obtain the lower bound for $m_{H^{++}}$ to be 42.9 GeV 
with the 95\% CL.

Because there is no study to directly search for the same-sign diboson signal at the LEP~II experiment, 
we here employ the results for the four charged lepton mode $e^+e^- \to Z^*/\gamma^*\to \ell^+\ell^+\ell^-\ell^-$~\cite{LEP}
in order to get bounds on $m_{H^{++}}$. 
The upper bound on the number of anomalous events can be applied to the case of the same-sign diboson decay scenario; 
i.e., $e^+e^-\to H^{++}H^{--}\to W^{+(*)}W^{+(*)}W^{-(*)}W^{-(*)}$. 
Because of the suppression of the signal cross section by the fourth power of the leptonic branching fraction of $W^\pm$, 
no substantial bound for $m_{H^{++}}$ can be obtained by using the data from Ref.~\cite{LEP}.

Consequently, the $\Gamma_Z$ measurement gives a bound $m_{H^{++}}>43$ GeV from the LEP I data.
On the other hand, no bound is obtained from the four charged lepton data at LEP II in our scenario. 
By carefully analyzing the hadronic events, a better bound could be obtained.


Let us consider whether the mass bound can be taken or not by using the current LHC data. 
The main production mode for $H^{\pm\pm}$ at the LHC is the pair production $pp\to \gamma^*/Z^* \to H^{++}H^{--}$ 
and the associated production $pp\to W^{\pm*} \to H^{\pm\pm}H^\mp$. 
The production cross sections for both the vector boson fusion $qQ\to q'Q'H^{\pm\pm}$ 
and the weak boson associated production $qQ\to W^{\pm *} \to H^{\pm\pm}W^\mp$ are proportional to 
$v_\Delta^2$, so that these modes are less significant because of $v_\Delta\ll v$. 
We note that there is the other production mechanism $qQ\to q'Q' H^{\pm\pm}A/H$ which is the unique process whose
difference of the electric charge between produced scalar bosons is two.
However, the cross section of this mode is small.

The signal processes are expected to be as follows: 
\begin{align}
pp &\to H^{++}H^{--}\to W^+W^+W^-W^-\to \ell^\pm\ell^\pm E_T\hspace{-4.5mm}/\hspace{2mm}+X, \notag\\
pp &\to H^{\pm\pm}H^{\mp}\to W^\pm W^\pm + X \to \ell^\pm\ell^\pm E_T\hspace{-4.5mm}/\hspace{2mm}+X \label{sig1}. 
\end{align}
The signal cross section before taking any kinematical cuts can be
estimated by 
\begin{align}
\sigma(\ell^\pm \ell^\pm E_T\hspace{-4.5mm}/\hspace{2mm}+X)&=
[\sigma(H^{++} H^{--})+\sigma( H^{\pm\pm} H^{\mp})]\notag\\
&\times\text{BR}(H^{\pm\pm}\to\ell^\pm\ell^\pm\nu\nu), \label{sig2}
\end{align}
where $\ell^\pm$ denote $e^\pm$ or $\mu^\pm$. 
The signal cross section of inclusive $H^{++}$ ($H^{--}$) production is
evaluated to be 0.72 (0.52)~pb at the leading order assuming
$m_{H^{\pm\pm}}=m_{H^\pm}=100$~GeV with the collision energy to be 7
TeV.
The branching ratio of the $H^{\pm\pm}$ decay can be calculated as a
square of that of $W^{\pm}$ for $m_{H^{\pm\pm}}>2m_W$.
However, it is no longer valid for $m_{H^{\pm\pm}}<2m_W$, because
at least one of the W bosons is off-shell. 
For $m_{H^{\pm\pm}}<2m_W$, the branching ratios of the same-flavor
dilepton decays are enhanced due to the interference of crossing
diagrams.

We here consider the bound for $m_{H^{++}}$ in the diboson decay scenario 
by using the results of the same-sign dilepton search reported by the ATLAS Collaboration~\cite{ATLAS} 
with the collision energy to be 7 TeV and the integrated luminosity to be 4.7 fb$^{-1}$. 
The 95\% CL upper limit $N_{95}$ for the event number 
of the process including the same-sign dilepton in the final states is listed in Ref.~\cite{ATLAS}. 
The limit is separately given into $e^\pm e^\pm$, $\mu^\pm \mu^\pm$, and $e^\pm\mu^\pm$ channels
after imposing the several choices of the invariant mass cuts for the same-sign dilepton system. 
The 95\% CL limit for the fiducial cross section 
$\sigma_{95}^{\text{fid}}$ is obtained by
\begin{align}
\sigma_{95}^{\text{fid}}=\frac{N_{95}}{\int \mathcal{L} dt \times \varepsilon_{\text{fid}}},
\end{align}
where $\varepsilon_{\text{fid}}$ is the efficiency for detecting events and $\int {\cal L} dt$ is the integrated luminosity of 4.7~fb$^{-1}$.
In Ref.~\cite{ATLAS}, 
the efficiency values are taken as 43\% for the $ee$ channel, 55\% for the $e\mu$ channel and 59\% for the $\mu\mu$ channel\footnote{
In fact, $\varepsilon_{\text{fid}}$ depends on 
the observed transverse momentum for charged leptons. According to Ref.~\cite{ATLAS}, $\varepsilon_{\text{fid}}$ for 
$e^\pm e^\pm$, $\mu^\pm \mu^\pm$, and $e^\pm\mu^\pm$ channels can be changed in the ranges of 43-65\%, 59-72\% and 55-70\%, respectively. 
In the derivation of $\sigma_{95}^{\text{fid}}$, $\varepsilon_{\text{fid}}$ is chosen to be the lowest value in Ref.~\cite{ATLAS}. }. 
We find that the data for the $\mu^+\mu^+$ channel with the invariant mass cut $M_{\ell\ell}>15$ GeV is the most effective 
choice to extract the more severe constraint on $m_{H^{++}}$. 
Therefore, we intend to bound $m_{H^{++}}$ by comparing $\sigma_{95}^{\text{fid}}$ for the $\mu^+\mu^+$ channel with the 
uncertainty of $\varepsilon_{\text{fid}}$ to the signal cross section.

In the numerical evaluation, 
$m_{H^+}$ is taken to be the same as $m_{H^{++}}$ to maximize the signal 
cross section. 
In order to generate the signal events, we use {\tt MadGraph5}~\cite{MG5} and the {\tt CTEQ6L} parton distribution functions~\cite{CTEQ}.
We impose the following basic cuts according to Ref.~\cite{ATLAS} for each muon: 
\begin{align}
&|\eta|<2.5,\quad p_T>20\text{ GeV},\label{basic}
\end{align}
where $\eta$ and $p_T$  are the pseudorapidity and the transverse momentum, respectively. 
Furthermore, in order to compare it to $\sigma_{95}^{\text{fid}}$, 
we impose the same invariant mass cut $M_{\ell\ell}>15\text{ GeV}$. 
In Table~\ref{cross}, the cross sections after imposing the basic cut and the invariant mass cut are shown. 
The signal cross section takes its maximum value at around $m_{H^{++}}=40$ GeV, because the cross section for the lower 
mass cases is much suppressed by the invariant mass cut. 
We also list the values for the decay width of $H^{\pm\pm}$ for $v_\Delta=1$ GeV, 
by which one can recognize that $H^{\pm\pm}$ can decay inside the detector.  

\begin{figure}[t]
\begin{center}
\includegraphics[width=80mm]{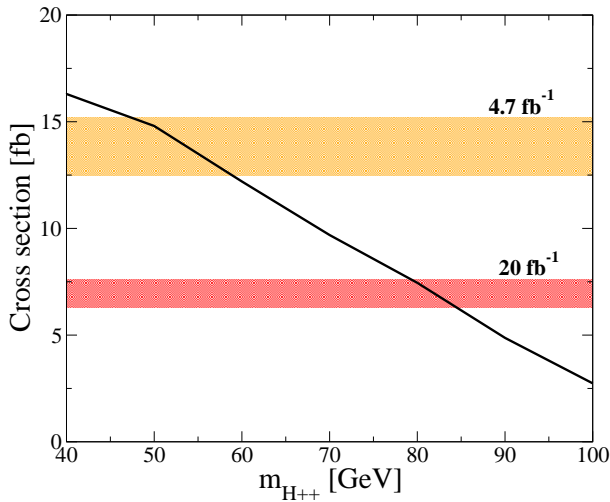}
\caption{The signal cross section after the $M_{\ell\ell}$ cut as a function of $m_{H^{++}}$ with the collision energy to be 7 TeV. 
The light (dark) shaded band shows the 95\% CL (expected) upper limit for the cross section from the data for the 
$\mu^+\mu^+$ channel with the integrate luminosity to be 4.7 fb$^{-1}$ (20 fb$^{-1}$). 
The width of the bands comes from the uncertainty of $\varepsilon_{\text{fid}}$ for the $\mu\mu$ system between 59\% and 72\%~\cite{ATLAS}. }
\label{fig2}
\end{center}
\end{figure}

In Fig.~\ref{fig2}, the same results are plotted as a function of $m_{H^{++}}$. The light shaded (orange colored) band 
gives the 95\% CL upper limit for the fiducial cross section with 4.7 fb$^{-1}$.
It can be seen that the HTM with the diboson decay scenario is excluded when $m_{H^{++}}$ is lower than about 60 GeV. 
We emphasize that this is the first substantial mass bound from the current LHC data, which is stronger than that obtained 
via the $\Gamma_Z$ data at the LEP experiment. 
The dark shaded (red colored) band is obtained by extrapolating the data to those for 20 fb$^{-1}$, where 
the cross section $\sigma_{95}^{\text{fid}}$ is expected to be smaller by a factor 2. 
This shows that the lower limit of $m_{H^{++}}$ becomes about 85 GeV with the accumulated integrated luminosity 20 fb$^{-1}$. 

Finally, we remark that the lower bound obtained in this Letter can be improved by the followings:
(i) the signal cross section is calculated at the leading order, which
can be modified by a factor of 1.2-1.3~\cite{Kfactor} with higher order corrections, 
(ii) although we have studied the same-sign dileptons only from the decay of $H^{\pm\pm}$,
they can also appear in the decay of $H^{\pm}$; e.g., $H^{\pm}\to W^{\pm}Z\to\ell^{\pm}\ell^+\ell^-\nu$.
However, the decay of $H^\pm$ strongly depends on the mass difference, so that
we here neglect these contributions as a conservative assumption for simplicity, 
(iii) a requirement of relatively hard jets in the same-sign dilepton events, like the analysis in Ref.~\cite{Nomura}, 
can enhance the significance of finding signal events from the background.
The mass bound could be improved by such an optimized analysis with the same data set.

In conclusion, 
we have studied the lower limit for $m_{H^{++}}$ in the diboson decay scenario 
where $H^{\pm\pm}\to W^{\pm(*)} W^{\pm(*)}$ is dominant by using the experimental data. 
By the LEP data for $\Gamma_Z$ the mass below 43 GeV is excluded with the 95\% CL, and 
the results from four charged lepton searches do not provide 
any significant constraint. 
We have found that a new lower bound is obtained to be about 60 GeV at the 95\% CL by using the current LHC data for 
the same-sign dilepton events with the collision energy to be 7 TeV and the integrated luminosity to be 4.7 fb$^{-1}$. 
This is the first substantial bound on $m_{H^{++}}$ from the LHC data. 
We have also shown by the extrapolation that the lower limit of $m_{H^{++}}$ becomes 
about 85 GeV with the accumulated integrated luminosity 20 fb$^{-1}$. 
When the LHC will rerun with the energy of 14 TeV, a more stronger bound can be obtained. 

The authors would like to thank Takaaki Nomura for useful discussions.
This work was supported in part by Grant-in-Aid for Scientific Research, Nos. 22244031, 
23104006 and 24340046, the National Science Council of R.O.C. under Grant No. NSC-101-2811-M-008-014, and 
the Sasakawa Scientific Research Grant from The Japan Science Society.


\end{document}